\documentclass
  [ams,pdfout]
	{br}
\RequirePackage[utf8]{inputenc}
\RequirePackage[english,brazil]{babel}
\RequirePackage{textcomp}
\RequirePackage{url}
\usepackage{graphics}
\usepackage{subfigure}
\usepackage{graphicx}
\usepackage{epsfig}
\usepackage{graphicx,indentfirst,amsmath,amsfonts,amssymb,amsthm,newlfont}
\usepackage{longtable}
\usepackage{cite}
\usepackage{hyperref}
\begin{document}

\begin{TitlePage}
	\Title{Linear Computer-Music through Sequences \\over Galois Fields}
	\RunningTitle{music over Galois Fields}
	\RunningAuthors{H. M. de Oliveira}
	\Author
		[hmo@de.ufpe.br]
		{H. M. de Oliveira}
		\Affil
			{%
				Federal University of Pernambuco, 
                Department of Statistics, Recife, Brazil
			}
		
	\Author
		[rcoliveira@uea.br]
		{R. C. de Oliveira}
        \Affil
        [A] 
        {Amazon State University, Department of Computer  Engineering, Manaus, Brazil
        }
		
	\Abstract{%
It is shown how binary sequences can be associated with automatic composition of monophonic pieces. We are concerned with the composition of e-music from finite field structures. The information at the input may be either random or  information from a black-and-white, grayscale or color picture. New e-compositions and music score are made available, including a new piece from the famous Lenna picture: the score of the e-music ``Between Lenna's eyes in C major.'' The corresponding stretch of music score are presented. Some particular structures, including clock arithmetic (mod 12), GF(7), GF(8), GF(13) and GF(17) are addressed. Further, multilevel block-codes are also used in a new approach of e-music composition, engendering a particular style as an ``e-composer.'' As an example, Pascal multilevel block codes recently introduced are handled to generate a new style of electronic music over GF(13).
	}%
\end{TitlePage}
\section{INTRODUCTION}

Many ways have been devised to compose music with aid of computers \cite{Minsky}, \cite{miranda2001composing}, \cite{Roads}, \cite{fucks1962mathematical}, \cite{manaris2003evolutionary}. One of the most common is adopting the basic principle of mapping some (binary or multilevel) data source to musical notes \cite{Rossing}. Among these, a random song is quite straightforward, simply by generating random sequences at the input of the mapping \cite{Niederreiter}, \cite{olson1961aid}. The input information can also come from another source such as $1/f$ noise \cite{voss19781}, or fractal structures \cite{su2007music}. In the case of a nucleotide sequence of a genome (or a particular genome stretch) of species \cite{Gena_Strom}, \cite{Gasser}, this is named DNA-music or gene-music \cite{pickover1995visualizing}, \cite{Munakata} (do not confuse with music generated by a genetic algorithm \cite{matic2010genetic}). In the same line of reasoning, an amino acid sequence in the generation of a protein can be used as data to generate the sound: the protein-music \cite{Dunn}, \cite{pickover1995visualizing}, \cite{King_Angus}, \cite{Takahashi_Miller}. Another approach may use an image as a data source to create a song (for example, \cite{Oberst}, \cite{Roads}). In this investigation, we propose miscellaneous of image-to-note maps from multilevel sequences over a Galois field to music notes, without the concern of put them in categories. DNA-music can be seen as sequences over GF(4), a particular case. We can also use nibbles from bytes (from a binary files, whatever be the information) to define both note and note value associated with each byte of the file. If you want to hear some interesting computer music generated by this approach, many sites are available, 
\begin{itemize}
\item \footnotesize{\url{https://www.youtube.com/watch?v=qNf9nzvnd1k}}
\item \footnotesize{{\url{http://www.toshima.ne.jp/\%7Eedogiku/TextTable/WhatisGM.html}}}
\item \footnotesize{{\url{http://www.genomamusic.com/genoma/ing/inicio.htm}}}
\item \footnotesize{{\url{http://larrylang.net/GenomeMusic/}}}
\end{itemize}
Indeed, many different mathematical-based descriptions are possible. There are other very interesting approaches to composing songs, including polyphonic \cite{lavrenko2003polyphonic}, much more sophisticated and attractive, such as the one developed at Sony Computer Science Laboratories, Paris \cite{Hadjeres_Pachet}. Not to mention \textbf{Iamus}, classical music's computer composer, that conceive the first complete album composed solely by a computer \cite{diaz2011composing}. Chermillier presented the Nzakara people's music from Congo and Sudan to five strings harp \cite{Chermillier}. Symmetries and group structure have long been exploited, as in \cite{Bailey}, \cite{Muzzulini}. Since all these approaches deal with sequences over finite field, this can be generalized. Furthermore, coded sequences (from multilevel error-correcting codes) can be used to replace the input sequences given rise some sort of signature of sequences, defining a ``style'' of musical composition. This is called here an e-composer (derived from the block code used to encoder the input sequence). In particular, we consider here the new (multilevel) block codes called Pascal Codes \cite{Paschoal}. Encoding the data from a given image, we generate a music composed by this virtual composer: in this case, Mr Pascal code (Equation~\ref{eq:MrPascal}).
\bigskip
\section{COMPUTER MUSIC FROM UNCODED BINARY SEQUENCES}
There are at least two straightforward ways to create e-music from particular input: i) music engendered from an image file ii) music derived from a random input sequence. In both case, we are initially concerned with 8-level information sequences. For picture-music, the image may be resized to a suitable size and then converted from color to black-white/grayscale as to yield note sequences in a standard octave-repeating (Diatonic scale). For random music generation, input are merely random sequences from a uniform numbers $X\sim$U(0,7) generator. From sequences generated over GF(7), a natural mapping is present in Table ~\ref{table:8-Gray}.
\begin{table*}[t!]
\centering
\caption{Musical notes codes as a function of the 8-Gray pixel level (or GF(8)).}
\begin{tabular}{c c c c c c c c c c}
\hline
signal &  0 & 1 & 2 & 3 & 4 & 5 & 6 & & over GF(7) \tabularnewline
8-gray & $\alpha^0$ & $\alpha^1$ & $\alpha^2$ & $\alpha^3$ & $\alpha^4$ & $\alpha^5$ & $\alpha^6$ & $\alpha^7$ &  over GF(8) \tabularnewline
C major & rest & C & D & E & F & G & A & B & \includegraphics[scale=0.15]
{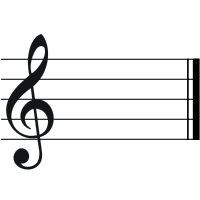}
\tabularnewline
F major & rest & F & G & A & Bb & C & D & E &  \includegraphics[scale=0.15]
{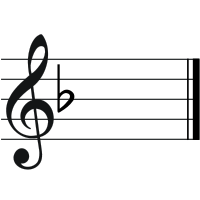} \tabularnewline
G major & rest & G & A & B & C & D & E & F\# &  \includegraphics[scale=0.15]
{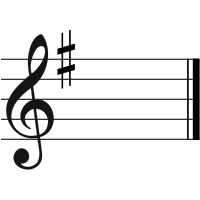} \tabularnewline
\hline
\end{tabular}
\label{table:8-Gray}
\end{table*}
Also, taking $\alpha$ as a primitive element of GF(8), different maps can be assigned using some key signature. For instance, when adopting F major as key signature, the mapping row 3 (Table ~\ref{table:8-Gray}) should be replaced by F G A Bb C D E F instead ($4^{th}$ row).\\
Why using the cumulative sum of note indexes? Reducing modulus 5 each note index may be not a good solution, since each single note always yields the same note value. As an option for defining the note value, its duration can also be computed by means of the accumulative sum of 8-level note indexes by reducing it modulus $q=p'=5$, where $q$ stands for an integer that establishes the finest quantization level of notes. If the finest level is assumed to be 1/16=1/$2^{(p'-1)}$, then $p'=5$. Smaller quantization is also possible, e.g. letting $p'=7$ we reach a 1/64. For instance, let us fists consider Table ~\ref{table:note_values} mapping.
\begin{table*}[ht!]
\centering
\caption{Note values defined in terms of the accumulative indexes by reducing it modulus $p'=5$.}
\begin{tabular}{c c c c}
\hline
symbol of GF(5) & duration & note & symbol \tabularnewline
\hline
0 & $1/2^0=1/1$ & Whole note (Semibreve) & \includegraphics[scale=0.15]
{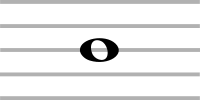} \tabularnewline
1 & $1/2^1=1/2$ & Half note (Minim) & \includegraphics[scale=0.15]
{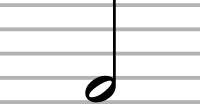} \tabularnewline
2 & $1/2^2=1/4$ & Quarter note (Crotchet) & \includegraphics[scale=0.15]
{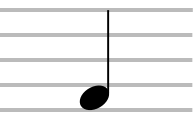} \tabularnewline
3 & $1/2^3=1/8$ & Eighth note (Quarver) & \includegraphics[scale=0.15]
{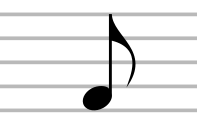} \tabularnewline
4 & $1/2^4=1/16$ & Sixteenth note (Semiquarver) & \includegraphics[scale=0.15]
{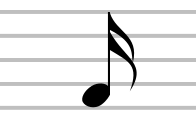}\tabularnewline
\hline
\end{tabular}
\label{table:note_values}
\end{table*}
\begin{figure}[b!]
\centering
\begin{subfigure}[{Between Lenna's eyes}]{\includegraphics[width=0.3\columnwidth]{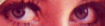} }
\end{subfigure}%
\begin{subfigure}[grayscale image]{\includegraphics[width=0.3\columnwidth]{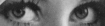} }
\end{subfigure}%
\begin{subfigure}[binarized image]
{\includegraphics[width=0.3\columnwidth]{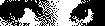} }
\end{subfigure}%
\caption{images from which the binarized file was used to compose the e-music piece.}
\label{fig:Lenna_eyes}
\end{figure}
In order to compose a short and naive e-music from the most used image of signal processing, the Lenna picture [\footnotesize{\url{https://en.wikipedia.org/wiki/Lenna}}\normalsize] was converted to black-and-white and just the region between Lenna's eye was cropped (Figure ~\ref{fig:Lenna_eyes}.) (562 bytes from which there are 416 data bytes, 72 pixel/inch, 105$\times$26 pixel). The file \footnotesize{\url{http://www.de.ufpe.br/~hmo/lennabetweeneyes.bmp}} \normalsize was read using \footnotesize{\url{http://www.onlinehexeditor.com}} \normalsize [note that the header field used to identify the bmp file is \textcolor{blue}{0$\times$42 0$\times$4d} [i.e. the character ``B'' then the character ``M'' in ASCII encoding]. The file size is $562_{10}$ bytes. The initial position of image data is $92_{16}$=$146_{10}$. Indeed, the score generated may or not contain the header information: in this case we throw out the header. Indeed, other image format can be used to compose e-music from the hex information. For instance, the gif file corresponding to the bmp  has 107 bytes (\footnotesize{\url{http://www.de.ufpe.br/~hmo/lennabetweeneyes.gif})\normalsize

\subsection{Variant 1}
The reading of any binary file in hexadecimal can be done in octal and each octal symbol set the note to play. In order to illustrate such a process, the first data line is (hex) \textcolor{blue}{7b b9 96 57 ee 95 b5 bf ff 53 88 00 00 00...}, which corresponds in (oct base) to the sequence \textcolor{blue}{173 271 226 127 356 225 265 277 377 123 210 0 0 0...}. It should be noted that the coding of a byte can be done in a different number of notes (between one and three notes). For example, the particular five bytes \textcolor{blue}{00 06 1b 27 e2} result in the following sequence of notes: \textcolor{blue}{ 
\{rest\} \{A\} \{E E\} \{F B\} \{E F D\}}, since 00=$0_8$ 06=$6_8$ 1b=$33_8$ 27=$47_8$ e2= $342_8$. Length: 828 musical notes.
\begin{center}
\footnotesize{
\textcolor{blue}{17327122612735622526527737712321000000354342334737311325}\\
\textcolor{blue}{337737737514200000336261212233771242452773773751020000027}\\
\textcolor{blue}{112427377377252133377377346276200000030615337377375326252}\\
\textcolor{blue}{377377153237300000017110127737734244151377372163773200000}\\
\textcolor{blue}{140657377356231333177376141573700000300325737733415226377}\\
\textcolor{blue}{370162736000002201023377363203231377374230735000000241377}\\
\textcolor{blue}{350144127277376120136000000013773603053177376003500000000}\\
\textcolor{blue}{377341203226377374003600000040017734226147377374003740000}\\
\textcolor{blue}{00017730020272377377003600000000177451127177377043600000}\\
\textcolor{blue}{000377026265377377043400000000375032157337377200034000000}\\
\textcolor{blue}{001700252311773773000000000000066177377377360000000000013}\\
\textcolor{blue}{452453773773700000000000125213637737737400000030000021251}\\
\textcolor{blue}{733773773740000003000001106273377377377100000003600001123}\\
\textcolor{blue}{115717737737733000000361400140302441473373773772450000037}\\
\textcolor{blue}{0125601212113773773773772101000000}}~~~~416 bytes of data\\
\end{center}
\subsection{Variant 2}
Another possible ``data bit-to-note code'' mapping is using nibbles (the first octal symbol) of the byte to set the note code and the second one to set the note value. Therefore, the initial sequence of the file, 
\textcolor{blue}{7b b9} $|$ \textcolor{blue}{96 57...} would be map as (173 mod 13) (271 mod 5) $|$ (226 mod 13) (127 mod 5) $|$ ... that is to say 4 1 $|$ 5 2 ...  The two first note are then \textcolor{blue}{Eb 1/2 E 1/4...}\\
\begin{figure*}[!t]
 \centering
 \includegraphics[scale=0.45]{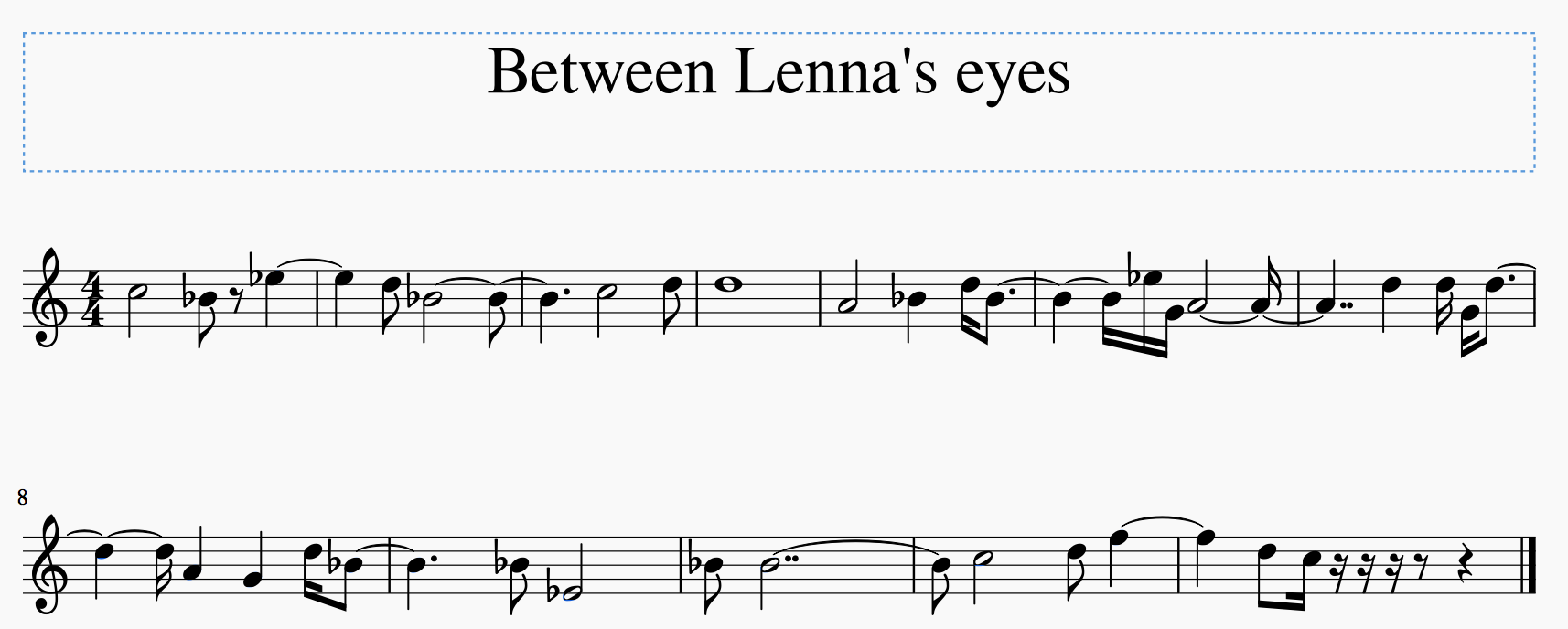}
 \caption{\small{Score of an excerpt of ``Between Lenna's eyes'' in C major.}}
 \label{fig:score_between_Lenna}
\end{figure*}
\subsection{Variant 3}
Further straightforward approach should be to use the Hamming weight \cite{Hamming} each data byte of the file to define the musical codes according with the Table ~\ref{table:Hamming_note}. The idea of cut out the code ``0'' becomes not only to fulfill the mapping of Table ~\ref{table:8-Gray} (from 8-point to 9-point), but also to avoid the common long sequences of zeros, which imply long rests. Here is the result for the file (Length: 237 musical notes):
\begin{table*}[!ht]
\centering
\caption{Musical notes codes for the Hamming weight of a byte.}
\begin{tabular}{c c c c c c c c c c}
\hline
level & 0 & 1 & 2 & 3 & 4 & 5 & 6 & 7 & 8 \tabularnewline
note  & erase & rest & C & D & E & F & G & A & B \tabularnewline
\hline
\end{tabular}
\label{table:Hamming_note}
\end{table*}
\footnotesize{64456457842544474588736433834787253488458856143787548856 25278424863832258646772652358543854342228634863342184357
72141842477484348641743586672158847135781480358813736781
34347824788415488514588621468862136888142467884512235788
4542321888821} (cf. Table ~\ref{table:Hamming_note})\\
\textcolor{blue}{GEEFGEFABECFEEEAEFBBADGEDDBDEABACFDEBBEFBB
FG rest EDABAFEBBFGCFCABECEBGDBDCCFBGEGA 
ACGFCDFBFEDBFEDECCCBGDEBGDDEC rest BEDFAAC rest
E rest BECEAAEBEDEBGE rest AEDFBGGAC rest FBBEA rest DFAB rest EBDFBB rest DAD
GAB rest DEDEABCEABBE rest FEBBF rest EFBBGC rest EGBBGC rest DGBBB rest ECE
GABBEF rest CCDFABBEFECDC rest BBBBC}.
\normalsize

\section{LINEAR COMPUTER MUSIC FROM SEQUENCES OF MULTILEVEL BLOCK CODES}
A possible way of choosing the note is to take into account each symbol of the GF(13)-valued codeword assuming a look-up table such as shown in Table ~\ref{table:temperate_GF13} to deal with the chromatic scale. Indeed, clock arithmetic (mod 12) can also be used, simply by neglecting rest. Again, The note values can be computed according to Table ~\ref{table:note_values}. Let us now pick a clef.
\begin{table*}[t!]
\centering
\caption{One-to-one mapping between notes of the temperate scale and symbols of GF(13).}
\begin{tabular}{c c c c c c c c c c c c c }
\hline
0 & 1 & 2 & 3 & 4 & 5 & 6 & 7 &  8  & 9 & 10 & 11 &12 \tabularnewline
\textcolor{blue}{rest} & \textcolor{blue}{C} & \textcolor{blue}{C\#} & \textcolor{blue}{D} & \textcolor{blue}{Eb} & \textcolor{blue}{E} & \textcolor{blue}{F} & \textcolor{blue}{F\#} & \textcolor{blue}{G} & \textcolor{blue}{G\#} & \textcolor{blue}{A} & \textcolor{blue}{Bb} & \textcolor{blue}{B} \tabularnewline
\hline
\end{tabular}
\label{table:temperate_GF13}
\end{table*}  
Here we also offer the use of multilevel error correcting codes over GF(13) to introduce some redundancy to the input data, i.e. by doing a block encoding on the sequence over GF(13). In order to play multilevel block codewords over GF(13), and starting new linear electronic music, an ($N,K$) code over GF(13) is chosen. Input: i) random generation of codewords as $k$ numbers with distribution $\sim$U(0,12), or ii) an image for the generation of codewords from the binary file of the picture and use as the information 
$\sum_1^{12} i$. A Pascal (13,8) blockcode over GF(13) has generator matrix $G$ given by (meet Mr Pascal):
\begin{equation}
\footnotesize{
\left| 
\begin{array}{c c c c c c c c c c c c c} 1&0&0&0&0&0&0&0&0&0&0&0&12\\
0&1&0&0&0&0&0&0&0&0&0&1&12\\
0&0&1&0&0&0&0&0&0&0&12&2&12\\
0&0&0&1&0&0&0&0&0&1&10&3&12\\
0&0&0&0&1&0&0&0&12&4&7&4&12\\
0&0&0&0&0&1&0&0&4&11&12&4&10\\
0&0&0&0&0&0&1&0&7&12&8&12&6\\
0&0&0&0&0&0&0&1&4&12&4&1&2\\ 
\end{array} \right|
\label{eq:MrPascal}}
\normalsize
\end{equation}
The Number of Distinct Excerpts is $13^K=815,730,721$. Coding the bmp image ``between Lenna's eyes'' with the generating matrix of Equation~\ref{eq:MrPascal}, we have the associated music generated by composer Mr Pascal. For instance, the simple repetition code (2,1) over GF(13) with random information symbols such as 11, 2, 5, 0... generate the sequence of repeated notes: {11 11 $|$ 2 2 $|$ 5 5 $|$ 0 0 ...} i.e.  
\textcolor{blue}{Bb 1/4 Bb 1/2 $|$ C\# 1/4 C\# 1/2 $|$ E 1/4 E 1/1 $|$ rest 1/1 rest 1/1 ...}. For a block code of length say $N$=10, a particular codeword (11 2 5 0 3 3 12 1 4 8) engender the following  accumulated (mod 3) sequence: 11 13 18 18 21 24 36 37 41 49
$\equiv$ 2 1 0 0 0 0 0 1 2 1. In this particular case, since $q$=3, only Whole note, Half note, and Quarter note are considered. The note sequence would thus be: \textcolor{blue}{Bb (1/4) C\# (1/2) E (1/1) rest (1/1) D (1/1) D (1/1) B (1/1) C (1/2) Eb (1/4) G (1/2)}\\
Now by assuming $q$=5 for the same codeword, the corresponding sequence is: 11 13 18 18	21 24 36 37 41 49 $\equiv$ 1 3 3 3 1 4 1 2 1 4\\
The note sequence this time would be:  
\begin{table*}[t!]
\centering
\caption{\small{Mapping GF(17) for the chromatic scale allowing different rest lengths (until Semiquaver-length rest).}}
\begin{tabular}
{c c c c c c c c c c c c c c c c c }
\tabularnewline
\hline
\footnotesize{0}&\footnotesize{1}&\footnotesize{2}&\footnotesize{3}&\footnotesize{4}& \footnotesize{5}&\footnotesize{6}&\footnotesize{7}&\footnotesize{8}&\footnotesize{9}& \footnotesize{10}&\footnotesize{11}&\footnotesize{12}&\footnotesize{13}&\footnotesize{14}&\footnotesize{15}&\footnotesize{16} \tabularnewline
\includegraphics[scale=0.4]{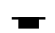} &
\includegraphics[scale=0.4]{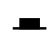} & \includegraphics[scale=0.25]{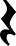} & \includegraphics[scale=0.22]{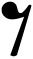} &
\includegraphics[scale=0.15]{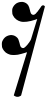}& \footnotesize{C}&\footnotesize{C\#}&\footnotesize{D}&\footnotesize{Eb}&\footnotesize{E}&\footnotesize{F}&\footnotesize{F\#}&\footnotesize{G}&\footnotesize{G\#}&\footnotesize{A}&\footnotesize{Bb}&\footnotesize{B}\tabularnewline
\tabularnewline
\hline
\end{tabular}
\label{table:temperate_GF17}
\end{table*}
\textcolor{blue}{Bb 1/2 C\# 1/8 E 1/8  rest 1/8 D 1/2  D 1/16  B 1/2 C 1/4 Eb 1/2  G 1/16}. Of course, many variants are possible. For example, to incorporate different lengths for rests, one can choose to use the structure on GF(17), using an (arbitrary) mapping of the type described in the Table \ref{table:temperate_GF17}.
\section{CONCLUSIONS}
\normalsize
A new approach to generating electronic music is presented which consists of the use of structures on finite fields, including multilevel block codes. Musical compositions employing block codes with different input symbols can be viewed as songs of the same composer (characterized by block code). Different variants are presented for both diatonic scale and chromatic scale. Some score of new musical pieces are available, including ``Between Lenna's eyes.''
%
%
%
\bibliographystyle{aes} 
\bibliography{bib} 

\begin{thebibliography}{10}

\bibitem{Minsky}
Marvin Minsky,
\newblock {\em ’’Music, Mind, and Meaning’’. In: Roads, Curtis (ed),
  The Music Machine: selected readings from the music journal},
\newblock The MIT Press, Cambridge, Mass., 1989.

\bibitem{miranda2001composing}
Eduardo Miranda,
\newblock {\em Composing music with computers},
\newblock CRC Press, 2001.

\bibitem{Roads}
Curtis Roads,
\newblock {\em The computer music tutorial},
\newblock MIT press, 1996.

\bibitem{fucks1962mathematical}
Wilhelm Fucks,
\newblock ``Mathematical analysis of formal structure of music,''
\newblock {\em IRE Transactions on Information Theory}, vol. 8, no. 5, pp.
  225--228, 1962.

\bibitem{manaris2003evolutionary}
Bill Manaris, Dallas Vaughan, Christopher Wagner, Juan Romero, and Robert~B
  Davis,
\newblock ``Evolutionary music and the zipf-mandelbrot law: Developing fitness
  functions for pleasant music,''
\newblock in {\em Workshops on Applications of Evolutionary Computation}.
  Springer, 2003, pp. 522--534.

\bibitem{Rossing}
Thomas Rossing, Richard Moore, and Paul Wheeler,
\newblock {\em The Science of Sound},
\newblock Addison Wesley, 3rd ed edition, 2002.

\bibitem{Niederreiter}
Harald Niederreiter,
\newblock {\em Random number generation and quasi-Monte Carlo methods},
\newblock SIAM, 1992.

\bibitem{olson1961aid}
Harry~F Olson and Herbert Belar,
\newblock ``Aid to music composition employing a random probability system,''
\newblock {\em The Journal of the Acoustical Society of America}, vol. 33, no.
  9, pp. 1163--1170, 1961.

\bibitem{voss19781}
Richard~F Voss and John Clarke,
\newblock ``’’1/f noise’’ in music: Music from 1/f noise,''
\newblock {\em The Journal of the Acoustical Society of America}, vol. 63, no.
  1, pp. 258--263, 1978.

\bibitem{su2007music}
Zhi-Yuan Su and Tzuyin Wu,
\newblock ``Music walk, fractal geometry in music,''
\newblock {\em Physica A: Statistical Mechanics and its Applications}, vol.
  380, pp. 418--428, 2007.

\bibitem{Gena_Strom}
Peter Gena and Charles Strom,
\newblock ``Musical synthesis of dna sequences,''
\newblock in {\em XI Colloquio di Informatica Musicale}, 1995, pp. 203--204.

\bibitem{Gasser}
Serge Morand and François Gasser,
\newblock {\em Langage des g\`enes - musique des hommes? Nodulations - la
  musique des g\`enes},
\newblock INRA Editions, Versailles, 1998.

\bibitem{pickover1995visualizing}
Nobuo Munakata and Kenshi Hayashi,
\newblock {\em Gene Music; Tonal assignments of Bases and Amino Acidsin:
  Visualizing Biological Information},
\newblock World Scientific, Singapore, 1995,
\newblock ISBN 9810214278.

\bibitem{Munakata}
Nobuo Munakata,
\newblock {\em Musical Representation of Gene Sequences, in:In Arts Medicine},
\newblock Saint Louis, 1997,
\newblock 0-918812-92-5.

\bibitem{matic2010genetic}
Dragan Mati{\'c},
\newblock ``A genetic algorithm for composing music,''
\newblock {\em Yugoslav Journal of Operations Research}, vol. 20, no. 1, pp.
  157--177, 2010.

\bibitem{Dunn}
John Dunn and Mary~Anne Clark,
\newblock ``Life music: the sonification of proteins,''
\newblock {\em Leonardo Music Journal}, vol. 32, no. 1, pp. 25--32, 1999.

\bibitem{King_Angus}
Ross~D King and Colin~G Angus,
\newblock ``Pm—protein music,''
\newblock {\em Computer applications in the biosciences: CABIOS}, vol. 12, no.
  3, pp. 251--252, 1996.

\bibitem{Takahashi_Miller}
Rie Takahashi and Jeffrey~H Miller,
\newblock ``Conversion of amino-acid sequence in proteins to classical music:
  search for auditory patterns,''
\newblock {\em Genome biology}, vol. 8, no. 5, pp. 405, 2007.

\bibitem{Oberst}
Thomas Oberst,
\newblock ``Blind graduate student ‘reads’ maps using cu software that
  converts color into sound,''
\newblock {\em Cornell Chronicle}, vol. 36, pp. 5, 2005.

\bibitem{lavrenko2003polyphonic}
Victor Lavrenko and Jeremy Pickens,
\newblock ``Polyphonic music modeling with random fields,''
\newblock in {\em Proceedings of the eleventh ACM international conference on
  Multimedia}. ACM, 2003, pp. 120--129.

\bibitem{Hadjeres_Pachet}
Ga{\"e}tan Hadjeres and Fran{\c{c}}ois Pachet,
\newblock ``Deepbach: a steerable model for bach chorales generation,''
\newblock {\em arXiv preprint arXiv:1612.01010}, 2016.

\bibitem{diaz2011composing}
Gustavo Diaz-Jerez,
\newblock ``Composing with melomics: Delving into the computational world for
  musical inspiration,''
\newblock {\em Leonardo Music Journal}, vol. 21, pp. 13--14, 2011.

\bibitem{Chermillier}
Marc Chemillier,
\newblock ``Math{\'e}matiques et musiques de tradition orale,''
\newblock {\em Math{\'e}matiques et sciences humaines}, , no. 178, pp. 11--40,
  2007.

\bibitem{Bailey}
Kathryn Bailey,
\newblock ``Symmetry as nemesis: Webern and the first movement of the concerto,
  opus 24,''
\newblock {\em Journal of Music Theory}, vol. 40 10.2307/843890, no. 2, pp.
  245--310, 1996.

\bibitem{Muzzulini}
Daniel Muzzulini,
\newblock ``Musical modulation by symmetries,''
\newblock {\em Journal of Music Theory}, vol. 39, no. 2, pp. 311--327, 1995,
\newblock 00222909.

\bibitem{Paschoal}
Arquimedes J~A Paschoal, Hélio~M de~Oliveira, and Ricardo~M Campello~de Souza,
\newblock ``A transformada numérica de pascal,''
\newblock in {\em Anais do XXXIII Simpósio Brasileiro de Telecomunicações},
  Juiz de Fora, 2015, SBrT.

\bibitem{Hamming}
Richard~W Hamming,
\newblock {\em Coding and Theory},
\newblock Prentice-Hall, 1980.

\end{thebibliography}
\end{document}